\documentclass[%
 reprint,
 amsmath,amssymb,
 prl,
 floatfix,
 groupedaddress,
]{revtex4-2}

\usepackage{graphicx}
\usepackage{dcolumn}
\usepackage{multirow}
\usepackage{bm}
\usepackage{bbold}
\usepackage{braket}

\begin{document}

%\preprint{APS/123-QED}

\title{Chiral electrons and spin selectivity at chiral-achiral interfaces}

\author{Xiaoming Wang}
\email{xiaoming.wang@utoledo.edu}
\affiliation{%
 Department of Physics and Astronomy and Wright Center for Photovoltaics Innovation and Commercialization, The University of Toledo, Toledo, Ohio 43606, USA
}%

\author{Yeming Xian}
%\email{xiaoming.wang@utoledo.edu}
\affiliation{%
 Department of Physics and Astronomy and Wright Center for Photovoltaics Innovation and Commercialization, The University of Toledo, Toledo, Ohio 43606, USA
}%

%\author{Peter C. Sercel}
%\email{}
%\affiliation{%
% Center for Hybrid Organic Inorganic Semiconductors for Energy, Golden, Colorado 80401, USA
%}%

\author{Yanfa Yan}%
\email{yanfa.yan@utoledo.edu}
\affiliation{%
 Department of Physics and Astronomy and Wright Center for Photovoltaics Innovation and Commercialization, The University of Toledo, Toledo, Ohio 43606, USA
}%

\date{\today}

\begin{abstract}
Chiral molecules can selectively transport electrons of a particular spin orientation, yet the underlying mechanism remain poorly understood. Here, we present theoretical evidence that electrons propagating through chiral materials with screw symmetry exhibit chirality themselves and carry symmetry-induced pseudo-angular momentum (PAM) which must be characterized by both spin and orbital components. The chiral electron transport is PAM polarized, with counterpropagating states exhibiting opposite PAM. At chiral-achiral interfaces, the PAM polarization is converted to spin polarization due to wave function matching, thereby establishing a novel principle for the chiral-induced spin selectivity. 
\end{abstract}

\maketitle

\section{I. Introduction}

Chiral-induced spin selectivity (CISS) is a phenomenon that chiral molecules can selectively transport electrons of a particular spin orientation~\cite{Ray1999,Gohler2011,Naaman2012,Waldeck2021}. 
It has emerged as a rapidly growing area of research with implications in enantioselective chemistry, long-range electron transfer, biorecognition, and spintronics~\cite{Naaman2019}.
In addition to molecules, the manifestations of CISS has been extended to semiconductors, e.g., chiral halide perovskites~\cite{Lu2019,Lu2020,Kim2021} which have demonstrated close-to-unity spin polarizations.  

Many theoretical models have been proposed to explain the CISS effect as manifested in the photoemission and transport experiments, but a consensus has not been reached~\cite{Evers2022}.
Most theories emphasize the importance of atomic spin-orbit coupling (SOC) as a key towards understanding the effect, since for time reversal-symmetric systems, SOC is considered as a natural source of spin-dependent phenomena.
It is generally adopted as a quantum transmission (scattering) treatment in the ballistic transport regime, either for bound~\cite{Guo2012} or unbound~\cite{Yeganeh2009} electrons, since the lengths of chiral molecules are only several nanometers which are smaller than the typical electron mean free path. 
Effective SOC of the chiral molecules has been considered to be a crucial part to induce the spin selectivity. 
However, the effective coupling strength of the chiral molecules is too small to account for the substantial magnitude of the observed spin polarization. 
In order to explain the experimental results, additional terms such as virtual bath~\cite{Guo2012,Matityahu2016,Volosniev2021}, nuclear effects~\cite{Wu2021,Fransson2021}, and band degeneracy~\cite{Dalum2019} together with SOC are included to enhance the spin polarization, yet the suitability of these terms is still under debate. 
Neglecting the SOC of chiral molecules, induced spin selectivity models~\cite{Gersten2013,Liu2021} that considered the potential selectivity of orbital polarization effect have been proposed, but the mechanism requires the assistance of the SOC of the substrate or electrode that contacts the molecule. Although strong SOC can be rationalized for heavy metal contacts such as Au~\cite{Gohler2011}, experiments have also been conducted on light-element contacts such as Al~\cite{Mishra2013} and graphite~\cite{Suda2019}. Nonetheless, current theories consider SOC as a necessary mechanism for CISS. 

While extensive manifestations of the CISS effect have been widely observed, uncovering the underlying mechanism remains a challenging endeavor. 
Here, we focus on the electronic band structure of the chiral materials, particularly the symmetry-induced quantum numbers of the wave functions for the transport electrons. 
We observe that chiral materials with screw symmetry manifest chiral electrons characterized by symmetry-induced pseudo-angular momentum (PAM), encompassing both spin and orbital components. 
The chiral electron transport is PAM polarized. 
The spin polarization is converted from the PAM polarization due to wave function matching at chiral-achiral interfaces, which reveals a novel mechanism of CISS.

\begin{figure*}[t]
\includegraphics[width=0.8\linewidth]{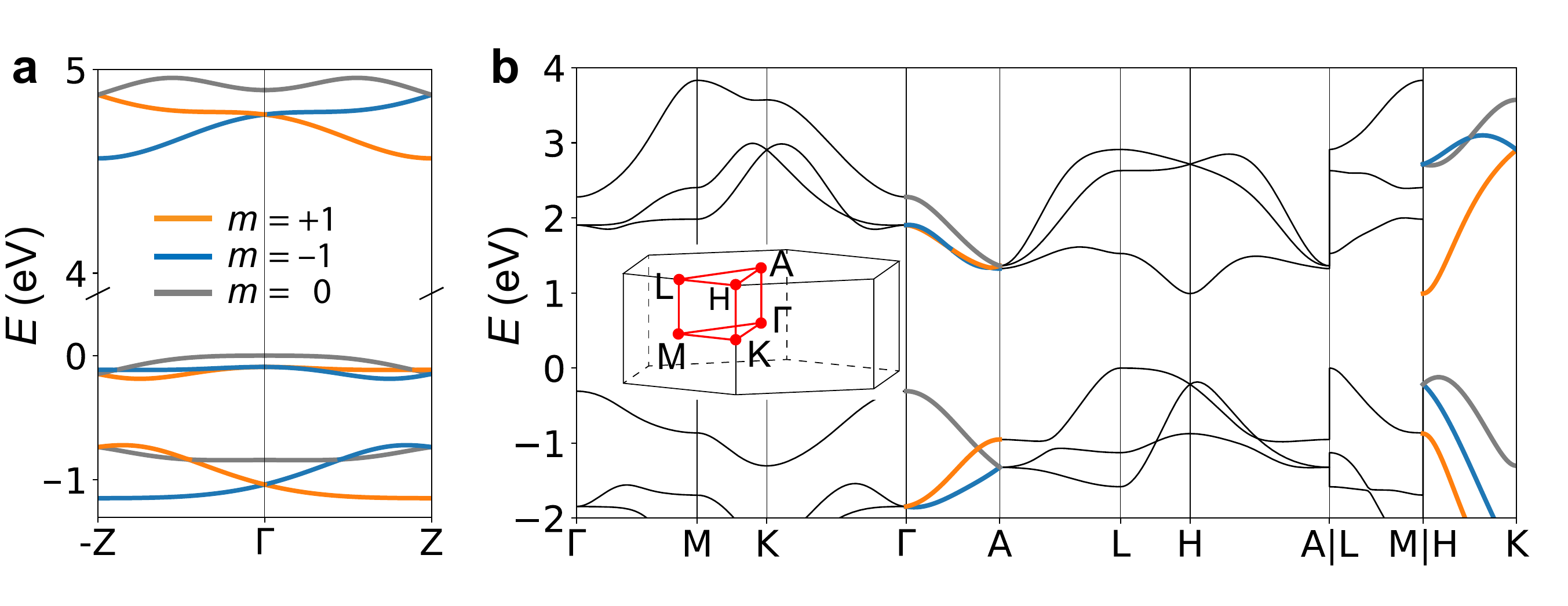}
\caption{\label{fig1} DFT band structures of (a) peptide $3_{10}$ helix, and (b) trigonal Se. The valence band maximum is set to 0. The PAMs are labeled and color coded. Inset of (b) shows the first BZ and high symmetry points of the hexagonal lattice.}
\end{figure*}

\section{II. Theory and results}

\subsection{A. Pseudo-angular momentum}

In atomic physics, magnetic quantum number is a quantum number used to distinguish quantum states of an electron according to its angular momentum along a given axis in space. 
The orbital magnetic quantum number ($m$) specifies the component of the orbital angular momentum (OAM) that lies along, usually, $z$ axis.
$m$ can be calculated from the OAM operator 
\begin{equation}
    \bm{L}_z \phi = m \hbar \phi
\end{equation}
with $\phi$ being the atomic orbital containing an azimuthal phase $e^{im\varphi}$. 
For Bloch orbitals, the OAM operator $\bm{L}_z$ no longer commutes with the Hamiltonian, however, similar azimuthal phase of the wave function arises due to the rotation symmetry. 
Hence, $m$ can be derived from the discrete $n$-fold rotational invariance~\cite{Yao2008,Cao2012,Zhang2015} 
\begin{equation}
    \bm{\mathcal{R}}\left(2\pi/n\right) \psi_k = \exp{(-i2m \pi/n)} \psi_k
\end{equation}
where $\bm{\mathcal{R}}(2\pi/n)$ is the rotation operator with the rotation angle $2\pi/n$ about the rotation axis $z$, and $\psi_k$ is the Bloch wave function with $k$ being the wave number. 
By this origin, $m$ is also called PAM and used for chiral phonons~\cite{Zhang2015}. 
Similarly, we call Bloch electrons with PAM chiral electrons.
The Bloch PAM is quantized and determined by the $n$-fold rotation symmetry $m=0,\pm 1, \pm 2, ..., \pm (n-1)$ mod $n$.

For chiral materials with screw axis $n_p$ (take $p=1$ for example)
\begin{equation}
    \label{eq3}
    \bm{\mathcal{S}}^n \psi_k = \bm{\mathcal{T}}(a) \psi_k = \exp{(-ika)} \psi_k
\end{equation}
where the screw operator $\bm{\mathcal{S}}=\{\bm{\mathcal{R}}(2\pi/n)|\bm{\mathcal{T}}(a/n)\}$ combines a rotation about the screw axis by angle $2\pi/n$ and a translation along the screw axis with vector $a/n$.
$a$ is the lattice constant.
From Eq.~\ref{eq3}, we have~\cite{Andreas2018}
\begin{equation}
    \label{eq4}
    \bm{\mathcal{S}}\psi_k=\exp{(-i2m\pi/n)} \exp{(-ika/n)} \psi_k
\end{equation}
Therefore, PAM can be derived from the screw operation for chiral materials with screw symmetry. 
Note, that $k$ is the quantum number derived from the translation symmetry (periodicity of $a$) with the first Brillouin zone (BZ) of $[-\pi/a,\pi/a]$.

\subsection{B. DFT band structures}

To show the chiral electrons with PAM, we demonstrate two examples by first-principles density functional theory (DFT) band structure calculations, as depicted in Fig.~\ref{fig1}. 
Our first example is a helical chain of right-handed peptide $3_{10}$ helix which is a typical secondary structure found in proteins and polypeptides. 
CISS with spin polarization of 60\% was reported for a single $\alpha$-helical peptide~\cite{Aragones2017}. 
The peptide $3_{10}$ helix has a $3_1$ screw axis.
The PAM of each electronic state is extracted from the eigenvalues~\cite{Gao2021} of the screw operator $\bm{\mathcal{S}}(3_1)$ according to Eq.~\ref{eq4}.
As shown in Fig.~\ref{fig1}a, the electronic bands are divided into groups with three entangled bands in each group, which is attributed to the nonsymmorphic rank of 3 as in the band topology~\cite{Parameswaran2013}. 
Each group possesses bands with $m$ of $0$, $-1$, and $+1$. 
There is collinear PAM-momentum locking for the bands with $m=\pm 1$, which indicates PAM polarized transport along the screw axis.
The OAMs of the bands of peptide were reported previously~\cite{Liu2021} with non-integer numbers since the OAM operator dose not commute with the Hamiltonian.
Instead, the PAMs derived from the screw operation can take only integers.

Above is a 1D example, as for 3D materials, we choose one of the simplest chiral semiconductors, namely, the elemental Se which crystallizes in space group $P3_121$ (right-handed) and $P3_221$ (left-handed). 
We focus on the former which has a $3_1$ screw axis along the crystallographic $c$ direction. 
The structure contains helical chains formed by selenium atoms through covalent bonding along $c$.
The chains are arranged in a hexagonal network with van der Waals interactions. 
The calculated band structure along high symmetry lines in the first BZ (inset) are shown in Fig.~\ref{fig1}b. 
The electronic states along $\Gamma$A and HK are invariant under $3_1$ screw operations, therefore, we calculate and label the PAMs for the bands along these two paths. 
Similar band topology can be seen from the band structure. 
Each band is identified with a specific PAM. 
Again, the chiral electron states are confirmed by our first-principles DFT calculations.

\subsection{C. Tight-binding formulation}

\begin{figure}[t]
\includegraphics[width=0.9\linewidth]{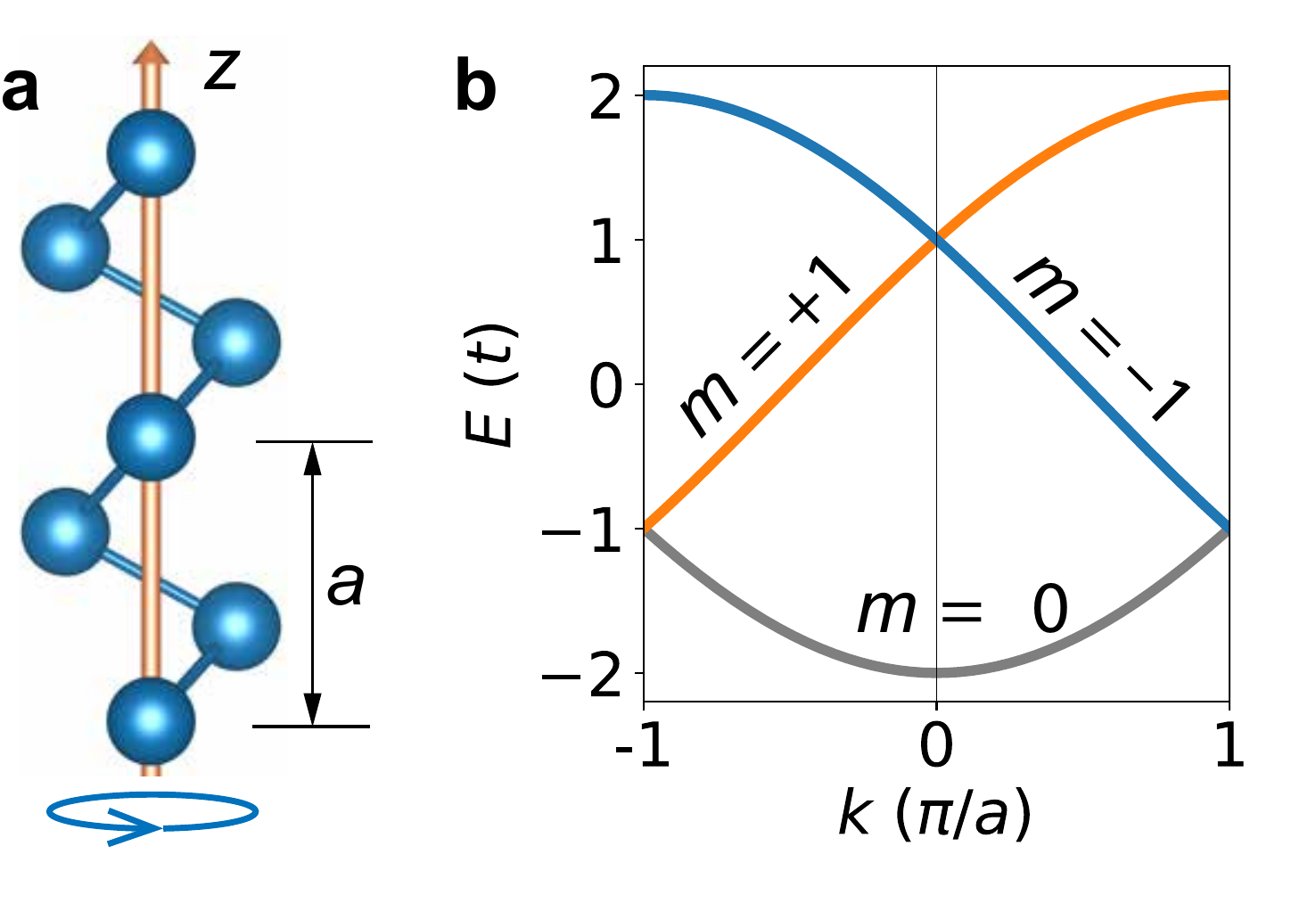}
\caption{\label{fig2} (a) Tight-binding model of a right-handed helical monoatomic chain with $3_1$ screw axis along $z$. $a$ is the translational lattice parameter. (b) The tight-binding band structure with PAM labeled. The hopping integral is $-t$.}
\end{figure}

The PAM of chiral electrons can be more clearly understood by tight-binding calculations.
We construct a tight-binding model composed of a helical monoatomic chain with $3_1$ screw axis along $z$, as shown in Fig.~\ref{fig2}a. 
Only one atomic orbital $\phi$ is considered on each atom.
The onsite energy is set to $0$.
The hopping integral between neighboring sites is $-t$ ($t > 0$). 
The pitch or translational periodicity along $z$ is $a$.

Considering the translation symmetry, the wave function of the helical chain is Bloch wave
\begin{equation}
    \label{eq5}
    \psi_k = \frac{1}{\sqrt{N}}\sum_{q=0}^{N} \sum_{l=0}^{2} \exp{[ik(q+l/3)a]} \phi_l
\end{equation}
Here, $N$ is the number of unit cells.
The calculated energy dispersion is 
\begin{equation}
    \label{eq6}
    E=
    \begin{cases}
        -2t\cos{(ka/3 - 2\pi/3)} \\
        -2t\cos{(ka/3)} \\
        -2t\cos{(ka/3 + 2\pi/3)} \\
    \end{cases}
\end{equation}
The three bands are shown in Fig.~\ref{fig2}b.
Acting the screw operation on the Bloch wave functions, we obtain for each band the PAM information which is labeled on the plot.
The three bands form a nonsymmorphic group that has PAMs of $0$, $-1$, and $+1$ with the bands of $m=\pm 1$ related to each other through time reversal symmetry.

If we consider the screw symmetry, Eq.~\ref{eq4} suggests a symmetry-adapted, generalized Bloch wave function~\cite{Izumida2009,Kashiwa2023} or helical wave
\begin{equation}
    \label{eq7}
    \psi_{k} = \frac{1}{\sqrt{N}} \sum_{l=0}^N  \exp{(ikla/3)} \exp{(i2ml\pi/3)} \phi_l
\end{equation}
$N$ is the total number of atoms. With this helical wave, the calculated energy dispersion is
\begin{equation}
    \label{eq8}
    E = -2t \cos{(ka/3 + 2m \pi/3)}
\end{equation}
$m$ is the screw-induced PAM which takes $0, \pm 1$ for $3_1$ screw symmetry. 
Inserting the $m$ values into Eq.~\ref{eq8}, we obtain exactly the same energy dispersion as Eq.~\ref{eq6}.
Therefore, from the helical basis, the PAM information is explicitly included in the wave form. 

\begin{figure}[t]
\includegraphics[width=0.8\linewidth]{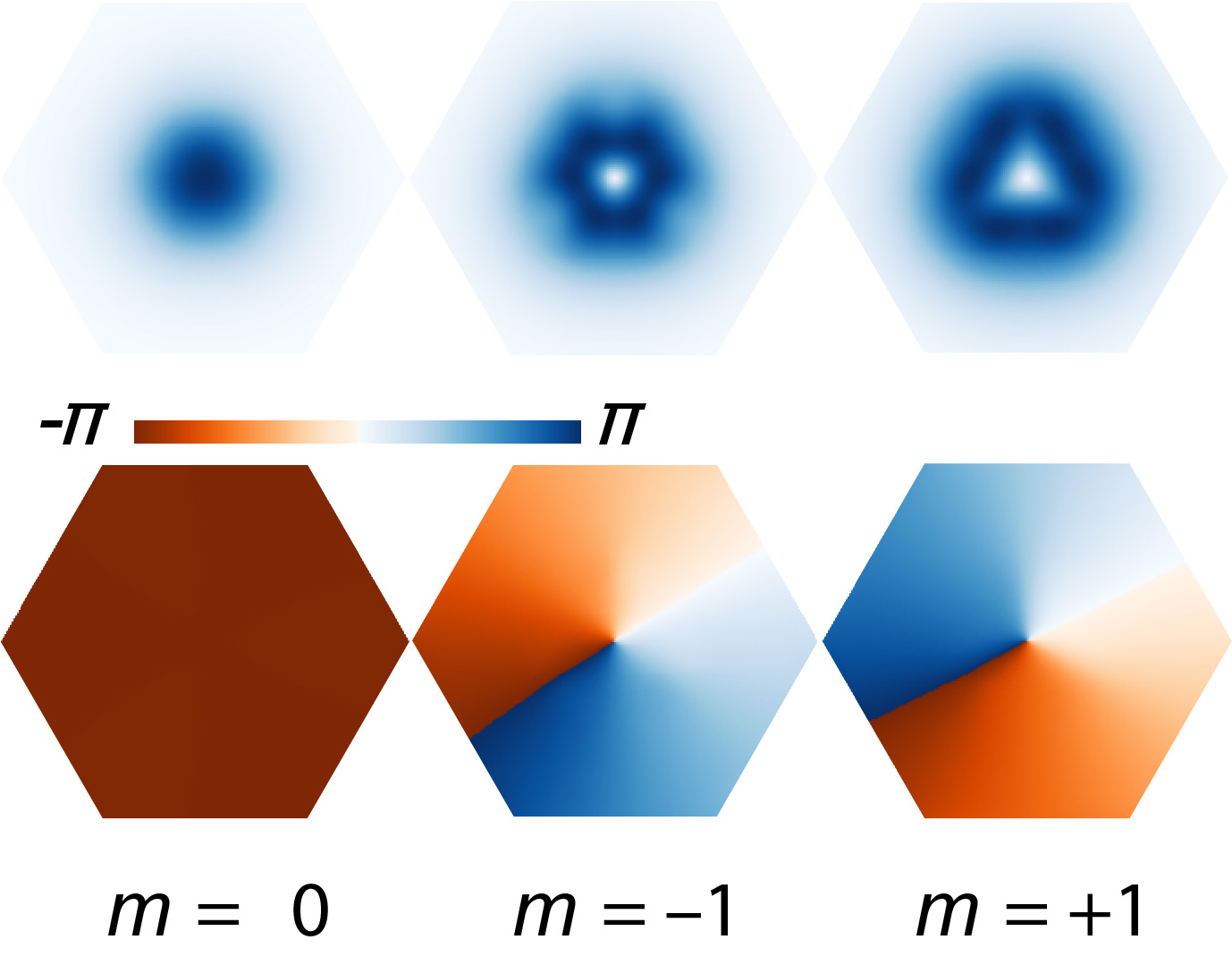}
\caption{\label{fig3}  Transversal wave amplitude
(first row) and phase (second row) distributions for the Bloch states with different PAMs.}
\end{figure}

Helical waves are well-known for vortex beams~\cite{Allen1992,Bliokh2007} for which the OAM is called topological charge that is defined~\cite{Bliokh2017,Lloyd2017} 
\begin{equation}
    m=\frac{1}{2\pi} \oint_{\mathcal{C}} \nabla \varphi \cdot \mathbf{u}
\end{equation}
where $\mathbf{u}$ is the unit vector tangential to the closed path $\mathcal{C}$ swirling the phase singularity. 
Fig.~\ref{fig3} shows the Bloch wave functions of the three states at $k=2\pi/3$ in Fig.~\ref{fig2}b, where we assumed $s$-type atomic orbital on each atom.
From the phase variations, the topological charge of each wave function can be clearly identified, which provides another way to calculate the PAM of Bloch electrons.

\begin{figure}[b]
\includegraphics[width=1.0\linewidth]{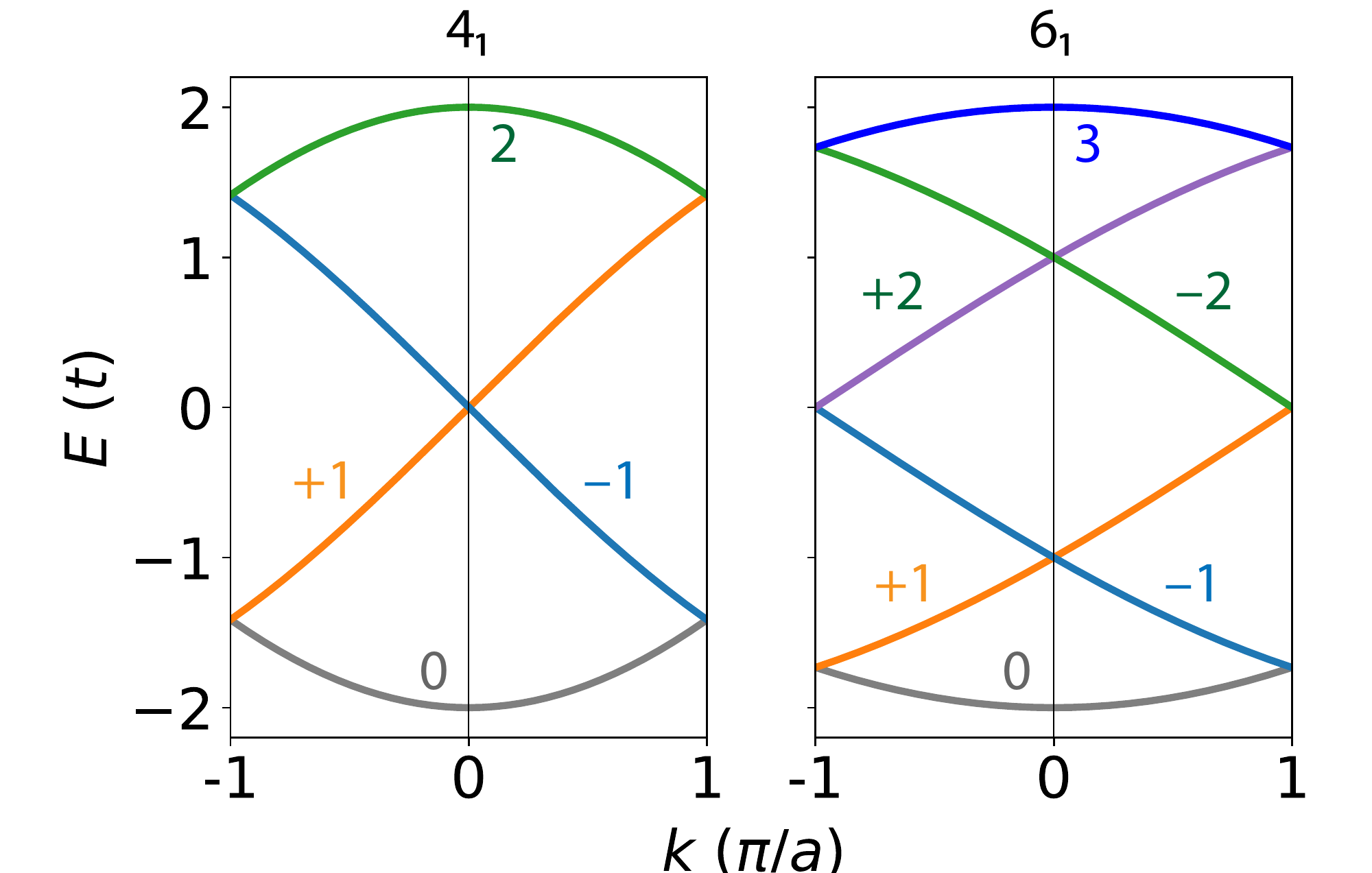}
\caption{\label{fig4}  Bloch bands for right-handed helical chains with $4_1$ and $6_1$ screw axes. PAMs are labeled.}
\end{figure}

Higher $m$ can be obtained for higher $n$-fold screw operations. The band structures of monoatomic helical chains of 4- and 6-fold screw axes are shown in Fig.~\ref{fig4}.
For even $n$, the PAM of $\pm n/2$ are equivalent.

\subsection{D. Chiral electrons with spin}

In previous discussions, we only take into account the orbital part of the electron wave functions.
As electron has an intrinsic angular momentum, spin, the PAM should also have a spin component.
Applying screw operation on spin wave function $\chi$, we have
\begin{equation}
    \bm{\mathcal{S}} \chi =\bm{\mathcal{R}} \left(2\pi/n\right)\chi = \exp{(-i2s\pi/n)} \chi
\end{equation}
where $s=\pm 1/2$ and $\chi = \ket{\uparrow}$ or $\ket{\downarrow}$. 
Note that the translation part of the screw operation does not affect the spin wave function.
Therefore, the symmetry-adapted helical wave function with spin included is~\cite{Izumida2009}
\begin{equation}
    \label{eq11}
    \Psi_{k} = \frac{1}{\sqrt{N}} \sum_{l=0}^N \exp{(ikla/3)} \exp{(i2jl\pi/3)} \phi_l \chi_l 
\end{equation}
where $j=m+s$ is the total PAM with both spin and orbital parts included.

It is important to distinguish the screw symmetry derived helical wave formulation from the translation symmetry derived Bloch wave formulation. 
For the latter, the spin wave function does not change under translations.
The total wave function is simply a direct product between the orbital part, i.e., Eq.\ref{eq5}, and the spin wave function $\chi$.
Note that translation is not a basic operation for the helical symmetry but the screw operation is.
For the spinless case, both of the formulations give the same band structure due to $\bm{\mathcal{S}}^n = \bm{\mathcal{T}}$. 
However, for spinfull case, $\bm{\mathcal{S}}^n \neq \bm{\mathcal{T}}$, the band structures will differ between the two formulations as we show below.

\begin{figure}[t]
\includegraphics[width=1.0\linewidth]{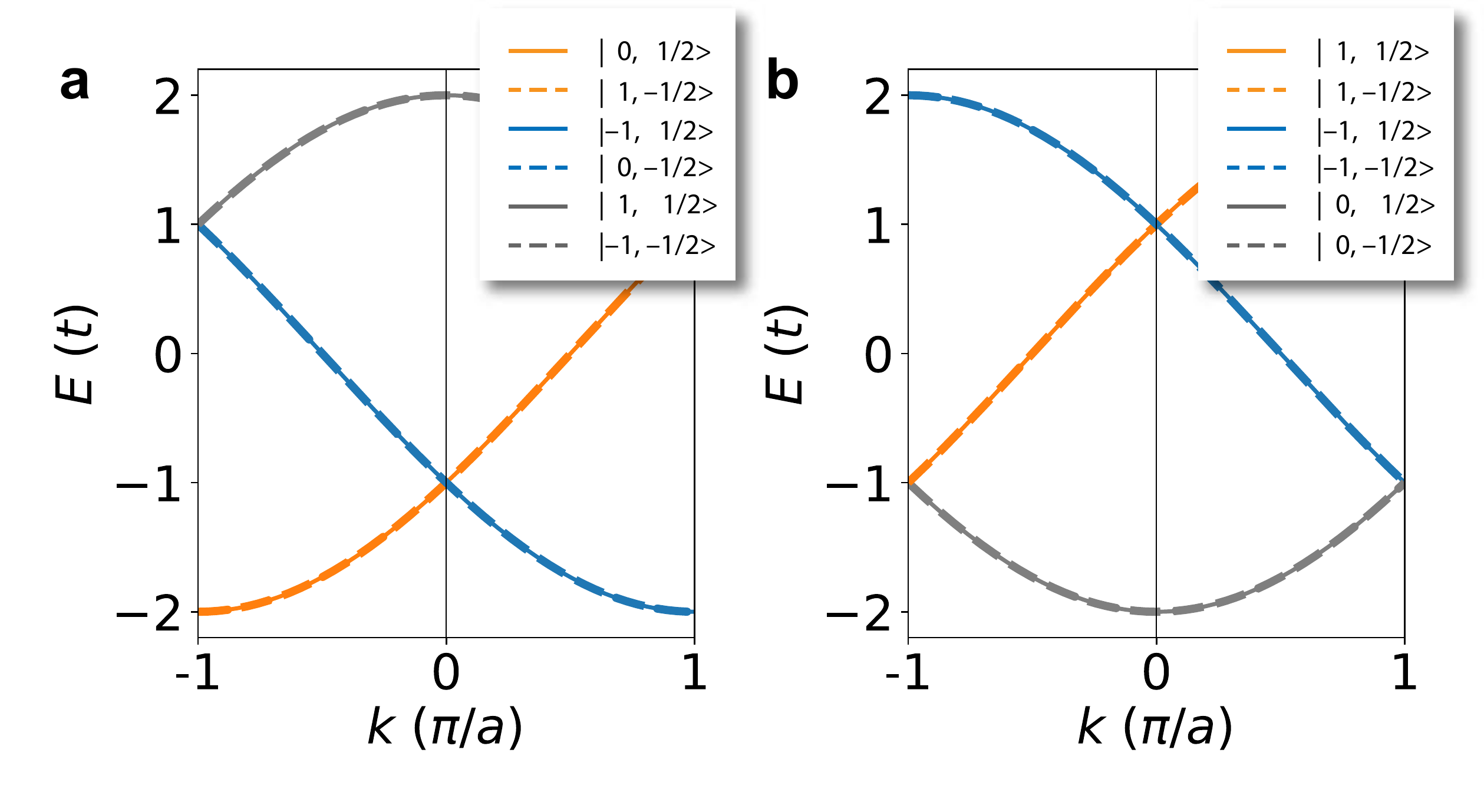}
\caption{\label{fig5}  Spinfull band structures of $3_1$ helical chain model obtained from the (a) helical and (b) Bloch formulations. Bands are labeled by the $\ket{m,s}$ notation. Solid and dashed lines denote spin up and down states, respectively.}
\end{figure}

Based on the wave function given in Eq.~\ref{eq11}, the calculated energy dispersion is (without SOC)
\begin{equation}
    \label{eq12}
    E=
    \begin{cases}
        -2t\cos{[ka/3 + 2(m+1/2) \pi/3]} \\
        -2t\cos{[ka/3 + 2(m-1/2) \pi/3]} 
    \end{cases}
\end{equation}
for the two spin channels.
Since there is no SOC, the two spin channels are degenerate.
The band dispersion are shown in Fig.~\ref{fig5}a.
From Eq.~\ref{eq12}, the bands can be classified by the total PAM $j$. 
The two spin channels with same $j$ have different $m$.
For example, the states with $j=1/2$ can be obtained with $m=0$ for the spin up channel $\ket{0, 1/2}$ (in the $\ket{m,s}$ notation) or $m=1$ for the spin down channel $\ket{1,-1/2}$, as denoted by the solid and dashed orange lines, respectively. 
Interestingly, in the energy range of $-2t \sim -t$, we have counter-propagating states with opposite $j$. 
Similar band features are known for quantum spin Hall states~\cite{Qi2011}. 
For the forward propagating electrons with $j=1/2$, the transport channels are $\ket{0,1/2}$ and $\ket{1,-1/2}$ which are degenerate. 
The signs of all the angular momentum quantum numbers are flipped for the backward direction due to time reversal symmetry. 
Therefore, the electron transport is PAM polarized.
From the energy dispersion equation, we can see that the sign of the polarization can be flipped for the hopping integral to be changed from $-t$ to $t$.

For comparison, we also show the band structure obtained from the Bloch formulation in Fig.~\ref{fig5}b.
The bands can be obtain from that of Fig.~\ref{fig2}b by simply adding the spin degeneracy. 
Therefore, in the Bloch formulation, the bands are classified by $m$ with each band representing two spin-degenerate states $\ket{m,\pm 1/2}$.

\begin{figure}[b]
\includegraphics[width=\linewidth]{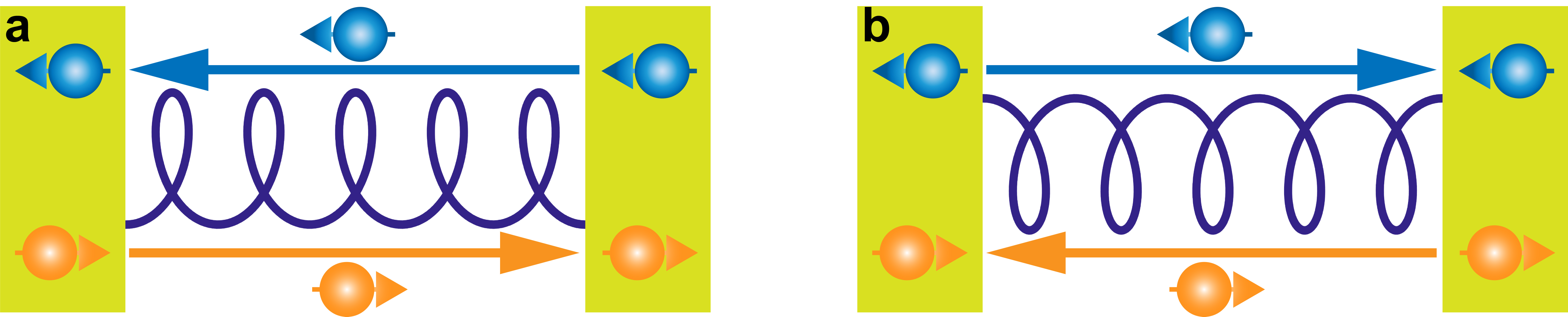}
\caption{\label{fig6} The spin selectivity at chiral-achiral interface. Schematic configurations of electron transport measurements for (a) right-handed and (b) left-handed chiral systems. The green rectangles denote metallic electrodes. The springs denote chiral materials with screw axis. Big orange and blue arrows indicate the transport directions. Balls with arrow denote electrons with spin. }
\end{figure}

\subsection{E. Chiral-induced spin selectivity}
To explain the spin selectivity, we adopt the band structure in Fig.~\ref{fig5}a with the energy range of $-2t \sim -t$ as an example and assume coherent transport.
From the above analysis of the band structure, the electron transport is PAM polarized with $j=1/2$ and $-1/2$ for the forward and backward propagation, respectively. 
For the forward direction, the two transport channels are denoted by $\ket{0,1/2}$ and $\ket{1,-1/2}$.
In the transport measurements (Fig.~\ref{fig6}), the chiral materials are connected with achiral electrodes, e.g., Au, Al, etc., within which the electron wave function is plane wave.
Therefore, the incoming electrons from the electrode are denoted by $\ket{0,\pm 1/2}$.
At the chiral-achiral interface, continuity condition requires that the wave functions of the two sides must match, for both orbital and spin parts. 
The wave function of the spin up electron $\ket{0,1/2}$ from the electrode matches the spin up transport channel with $m=0$, whereas the spin down electron has an orbital mismatch between plane wave and helical wave ($m=1$).
Therefore, only the spin up electrons can be transmitted through the achiral-chiral interface.
Reversing the transport direction, the transport channels are changed to $\ket{0,-1/2}$ and $\ket{-1,1/2}$, which only allow the spin down electrons from the electrode to get through, as shown in Fig.~\ref{fig6}a.
Therefore, the PAM polarization is converted to spin polarization due to wave function matching at the chiral-achiral interface.  
Similar matching applied for the second chiral-achiral interface, the outcoming electron to the right electrode keeps the state $\ket{0,\pm 1/2}$. 
For the typical photoemission experiments~\cite{Gohler2011}, the second interface is chiral-vacuum, the emitted electrons still preserves plane wave feature in the vacuum. 
This wave function matching results in a PAM conservation mechanism at the chiral-achiral interface.
Changing the chirality of the chiral materials, the angular momentum quantum numbers ($j,m,s$) of the transport channels flip sign due to time reversal operation, which leads to opposite spin polarizations (Fig.~\ref{fig6}b).

As for systems other than $3_1$ symmetry, the electron transport across the chiral-achiral interface follows the selection rule $\Delta j=nq/2$ for chiral materials with $n$-fold screw symmetry, where the $\Delta j$ is the difference of the PAM between the chiral and achiral sides, and $q$ is an integer. 
Therefore, electron transport is forbidden across the chiral-achiral interface for the chiral bands with $j$ of $\pm 3/2$ as in $6_1$ screw symmetry. 
However, extra PAM may be introduced by chiral phonons~\cite{Zhang2015,Ishito2022} to achieve the PAM conservation.

\section{III. Conclusion}
In conclusion, we revealed the mechanism of CISS, without calling for SOC, by investigating the screw symmetry-induced PAM of the chiral electrons propagating through the chiral materials. 
Spin selectivity is achieved at chiral-achiral interface by imposing the continuity condition that matches both the orbital and spin parts of the wave functions from both sides. 
Our findings provide profound insights into the physics of the CISS phenomena and may advance the CISS based applications and materials design.

\section{acknowledgments}
This work was supported as part of the Center for Hybrid Organic Inorganic Semiconductors for Energy (CHOISE) an Energy Frontier Research Center funded by the Office of Basic Energy Sciences, Office of Science within the U.S. Department of Energy. 
The first-principles calculations were supported by the National Science Foundation under contract number DMR-1807818 and performed using computational resources sponsored by the Department of Energy's Office of Energy Efficiency and Renewable Energy and located at the National Renewable Energy Laboratory, and the resources of the National Energy Research Scientific Computing Center (NERSC), a US Department of Energy Office of Science User Facility located at Lawrence Berkeley National Laboratory, operated under contract DE-AC02-05CH11231 using NERSC award BES-ERCAP0023945.  
X.W. thanks Peter C. Sercel for helpful discussions and Binghai Yan for providing the POSCAR file of the peptide helix.

\appendix

\section{APPENDIX}
%\begin{figure}[htbb]
%\includegraphics[width=\linewidth]{fig-2.pdf}
%\caption{\label{figa} Band structures of (a) peptide $3_{10}$ helix, right-handed Se with (b) unit cell and (c) $9 \times 9 \times 1$ supercell. The OAMs are color coded and labeled. The valence band maximum is set 0.}
%\end{figure}
%\section{APPENDIX B. Methods}
%Tight-binding calculations were performed using the PythTB package.
First-principles band structure calculations were performed using the Vienna Ab initio Simulation Package~\cite{Kresse1996-1,Kresse1996-2} with projector augmented-wave potentials~\cite{Blochl1994}. 
A kinetic energy cut-off of 520 eV was used to expand the wave functions. 
Perdew–Burke-Ernzerhof~\cite{Perdew1996} functional was employed for the exchange-correlation interaction. 
The Brillouin zone was sampled with $\Gamma$-centered $1\times 1 \times 6$ (peptide), $8 \times 8 \times 8$ (Se) k-meshes, respectively. 
Eigenvalues of the screw operators were calculated using the irvsp code~\cite{Gao2021}.

%\nocite{*}
\email{xiaoming.wang}
%\bibliography{refs}% Produces the bibliography via BibTeX.
\providecommand{\noopsort}[1]{}\providecommand{\singleletter}[1]{#1}%

\bibliographystyle{apsrev4-2}

\end{document}